\documentstyle[prl,aps,psfig]{revtex}
\begin{document}
\draft

\twocolumn[\hsize\textwidth\columnwidth\hsize\csname@twocolumnfalse\endcsname

\author{Dario Alf\`e}

\address{ Department of Earth Sciences, and Department of Physics and
Astronomy, \\ University College London, Gower Street, London, WC1E
6BT, U.K }

\title{The melting curve of MgO from first principles simulations}

\maketitle

\begin{abstract}
First principles calculations based on density functional theory, both
with the local density approximation (LDA) and with generalised
gradient corrections (GGA), and the projector augmented wave method,
have been used to simulate solid and liquid MgO in direct coexistence
in the range of pressure $0\le p \le 135$~GPa. 
The calculated LDA zero pressure melting temperature is $ T_m^{\rm
LDA} = 3110 \pm 50 $~K, in good agreement with the experimental
data. The calculated GGA zero pressure melting temperature $ T_m^{\rm
GGA} = 2575 \pm 100 $~K is significantly lower than the LDA one, but
the difference between the GGA and the LDA melting curves is greatly
reduced at high pressure. The LDA calculated zero pressure melting
slope is $dT/dp \sim 100$~K/GPa, which is more than three times higher
than the currently available experimental one ((Zerr and Boehler,
Nature {\bf 371}, 506 (1994)). As the pressure is increased, the
melting curve deviates significantly from the experimental one. At the
core mantle boundary pressure of 135~GPa MgO melts at $T_m = 8140 \pm
150$~K.
\end{abstract}
]

MgO (periclase) is a very important material for a number of reasons.
At ambient conditions it has the rock-salt structure of NaCl, and has
the peculiarity of not showing any phase transition at least up to 227
GPa~\cite{duffy95}. For this reason it is often used as a pressure
medium in high pressure solid media devices.  The periclase
related structure Mg(Fe)O-magnesiow\"ustite is believed to be present
in large quantities in the Earth's lower mantle, therefore knowledge of the
behaviour of MgO under pressure is very important for our
understanding of the lower mantle. In particular, the melting behaviour of
periclase under pressure is important to put constraints on the
solidus in the lower mantle:
a low melting temperature of periclase may point to a low eutectic point,
and in particular may support the
suggestions of the presence of partial melt in the 'ultra-low velocity
zone'~\cite{williams96}.

The only available experiment for the melting behaviour of MgO under
pressure has been performed by Zerr and Boehler~\cite{zerr94} (ZB),
who measured the melting curve of MgO up to 32 GPa. They found a slope
of the melting curve at zero pressure $dT_m/dp \sim 30 $~K/GPa, and a
relatively low extrapolated melting temperature at lower mantle
pressures; a result which may support the presence of partial melt at
the bottom of the mantle.

A number of theoretical calculations, based on empirical potentials,
have attempted to determine the melting curve of periclase up to lower
mantle
pressures~\cite{vocadlo96,cohen94,belonoshko96,strachan01,tangney04}. The
results of these calculations are somewhat scattered, but they all
predict significantly higher than experiments zero pressure melting
slopes, and consequently much higher melting temperatures of periclase
at lower mantle pressures.  

Here I report first principles
calculations of the whole melting curve of MgO in the pressure range
0-135~GPa. Melting points have been calculated by performing direct
first principles simulations of solid and liquid MgO in coexistence.
Since a large number of atoms is needed to represent correctly solid
and liquid in equilibrium, the method is extremely computationally
intensive, however, the feasibility of the coexistence approach within
first principles calculations has been recently
demonstrated~\cite{alfe03,ogitsu03,bonev04}.  In the constant volume
constant internal energy ($NVE$ ensemble) approach to the coexistence
method it has been shown that liquid and solid can coexist for long
times, provided $V$ and $E$ are appropriately
chosen~\cite{alfe03,morris94,alfe02b}.  The average value of the
pressure $p$ and temperature $T$ over the coexisting period then give
a point on the melting curve.  Size effects have been studied quite
extensively, and it was shown that correct results, including in MgO,
can be obtained in systems containing more than 500
atoms~\cite{belonoshko96,alfe03,belonoshko00b,morris02}, although
recent calculations on LiH have also been performed on system
containing 432 atoms~\cite{ogitsu03}.

The zero pressure crystal structure of MgO is the same as the NaCl
structure, and since no transitions have been experimentally observed
up to at least 227~Gpa~\cite{duffy95}, I assumed that melting occurs
from this structure. However, I note that at high temperature this is
not necessarily true, and melting may occour from a different crystal
structure, as recently suggested by Aguado and Madden~\cite{aguado04}.
The existence of a more stable solid phase would increase the melting
temperature.  

A possible source of error in the calculations may be due to the
neglecting of the formation of defects in the solid. Although defects
formation is not explicitly excluded in the current approach, it is
unlikely that the simulations are long enough to allow for significant
ionic diffusion in the solid. A concentration $c$ of defects would be
responsible for a decrease of the solid Gibbs free energy per atom
$g_{\rm v} \sim -k_{\rm B}Tc$~\cite{dewijs98}, and a consequent
increase in the melting temperature $\delta T_m \sim g_{\rm v}/s_m$,
where $s_m$ is the entropy of melting. The concentration $c$ of
defects in MgO near the melting temperature is given by $c =
\exp\{{-E_{\rm v}/2k_{\rm B}}T\}$, where $E_{\rm v}$ is the formation
energy of the defect. An estimate of $c$ comes from the value of the
formation energy of a Schottky defect, which in MgO has a value
between 4 and 7~eV~\cite{mackrodt82}. Recent density functional theory 
(DFT)~\cite{devita92,alfe05} and quantum
Monte Carlo~\cite{alfe05} calculations point to a value close to 
7~eV, but even using the
lower value 4~eV, and $T = 3000$~K for the melting temperature, we
obtain $c \sim 6 \times 10^{-4}$, which results in a completely
negligible correction to the melting temperature.

Present calculations have been performed using density functional
theory with various approximations for the exchange-correlation (XC)
functional. I used the VASP code~\cite{kresse96}, with the
implementation of an efficient extrapolation of the charge
density~\cite{alfe99e}, and the projector augmented wave (PAW)
method~\cite{blochl94,kresse99}. Single particle orbitals have been
expanded in plane-waves, with a plane-wave cutoff of 400~eV. The Mg
potential has a core radius of 1.06~\AA, and the $3s^2$ electrons in
valence; the O potential has a core radius of 0.8~\AA~and the $2s^2
2p^4$ electrons in valence. The structural properties of MgO in its
rock-salt zero pressure crystal structure are compared with the
experimental results in Table~\ref{tab:struct}. Pressure against volume curves are compared
with experiments in the inset of Fig.~1. In Table~\ref{tab:struct} I
also report the value of the transition pressure between the rock-salt
and the CsCl structures, which is in the region of 500 GPa.  I also
tested a "small core" Mg potential with both $3s^2$ and $2p^6$
electrons in valence. Results are reported in Table~\ref{tab:struct}.
The zero temperature MgO crystal has a band gap of several eV's, but
this gap is significantly reduced at high temperature, where electron
excitations become important, therefore finite temperature
calculations have been performed by minimising the Mermin
functional~\cite{mermin65}.  In fact, the electronic entropy
contribution to the free energy of the system is non-negligible, and
at zero pressure it lowers the free energy of the liquid with respect
to the free energy of the solid by almost 0.1 eV. Given a zero
pressure entropy of melting is $\sim 2.2~k_{\rm B}$/atom (see
Tab~\ref{tab:results}), the electronic entropy is responsible for a
lowering of the zero pressure melting temperature of $\sim 500$~K.

The coexistence simulations have been performed using the local
density approximation (LDA) for the XC functional, and the Mg
potential with only $2s^2$ in valence, using the $NVE$ ensemble for
systems containing 432 atoms ($3\times 3\times 6$ cubic supercell).
The time step was 1 fs and the convergency threshold on the total
energy $2\times 10^{-8}$~eV/atom. With these prescriptions the drift
in the micro-canonical total energy was less than $0.5$~K/ps.
Simulations were carried out for up to 20 ps, and performed using the
$\Gamma$ point only. Spot checks with
Monkhorst-Pack~\cite{monkhorst76} ($2\times 2\times 1$) and ($2\times
2\times 2$) ${\bf k}$-point grids showed energies converged better
than 0.1~meV/atom and pressure converged better than 1~MPa.  A
correction term of 3.7 (5.7) GPa due to the lack of convergency with
respect to the plane wave cutoff has been added to the calculated
pressures in the low (high) pressure regions.

To prepare the system I have followed the same procedure employed in
Refs.~\cite{alfe03,alfe02b}. A perfect crystal is initially
thermalized to a guessed melting temperature $T_{\em guess}$, then the
simulation is stopped, half of the atoms are clamped and the other
half are freely evolved at very high temperature until melting occurs,
then the liquid is thermalized back at $T_{\em guess}$. At this point
the system is being freely evolved in the $NVE$ ensemble.  Note that
in $(V,E)$ space melting is represented by a ``band'', and not by a
line as in $(p,T)$ space. In particular, for every fixed $V$, a whole
piece of melting curve can be evaluated by varying $E$ appropriately.
The amount of total energy $E$ given to the system can be tuned by
assigning different initial values to the
velocities~\cite{velocity}. Now, unless the value of $E$ is in the
right range for the volume $V$, the system will either completely melt
or completely solidify. For MgO this happens very quickly, typically
in less than 1 ps. Therefore, for each fixed $V$, a certain number of
``trial and errors'' steps are required in order to find the right
value of $E$, for which solid and liquid coexist for long time.

\begin{figure}
\centerline{\psfig{figure=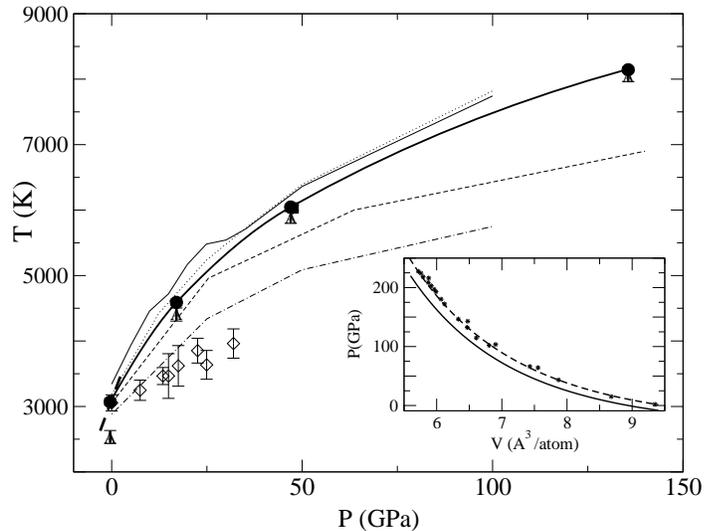,height=3.2in,angle=-90}}
\caption{Melting curve of MgO obtained with present DFT-LDA
coexistence simulations performed on 432-atom cells (black dots and
heavy solid line), 1024-atom cell (black square), and present DFT-GGA
results (triangles), compared with experiments (open
diamonds)~\protect\cite{zerr94} and theoretically determined curves
based on empirical potentials: Belonoshko and Dubrovinski (chained
line)\protect\cite{belonoshko96}, Strachan et al. (dashed
line)~\protect\cite{strachan01}, Cohen and Zong (dotted
line)~\protect\cite{cohen94}, Vo\v{c}adlo and Price (solid
line)~\protect\cite{vocadlo96}, Tangney and Scandolo (heavy dashed
line)~\protect\cite{tangney04}. Inset: pressure as function of
atomic volume for rock-salt MgO calculated with DFT-LDA (solid line) and DFT-GGA (dashed line),
and compared with experiments (stars)~\protect\cite{duffy95}. Calculations do not include zero point
motion. }\label{fig:melting}
\end{figure}

Simulations have been performed at a number of points in the pressure
range 0-135~GPa.  The points on the melting curve obtained from these
simulations are displayed in Fig~\ref{fig:melting}, and compared with
the experimental results of ZB, and with previously calculated melting
curves~\cite{vocadlo96,cohen94,belonoshko96,strachan01,tangney04}.
The solid curve between points has been obtained by interpolating with
third order polynomials, with the conditions for the coefficients
given by the coordinates and the slopes at the points.  The slope of
the melting curve is obtained from the Clausius-Clapeyron relation $
dT / dp = v_m / s_m$, where $v_m$ and $s_m$ are the volume and the
entropy change on melting respectively. For a chosen point on the
melting curve $(p_m,T_m)$, $v_m$ is calculated from independent
simulations on solid and liquid MgO, with the respective volumes
adjusted until the calculated pressures are equal (within $\sim
0.5$~GPa) to the chosen value $p_m$. From these simulations it is also
possible to obtain $s_m$, given by $s_m = (e_m + p_m v_m )/T_m$,
%~\cite{errors}, 
where $e_m = e_{\rm liq} - e_{\rm sol}$, with $e_{\rm liq}$, $e_{\rm
sol}$ the internal energies in the liquid and the solid simulations
respectively. These simulations have been performed on cells
containing 64 atoms, for over 40 ps, and spot checked with simulations
performed with 216 atoms, which showed essentially no difference in
$e_m$ within a statistical error of $\sim 10$ meV/atom. Pressures
calculated with the 64-atom and 216-atom cells differed by less than
0.5 GPa.  Results for $v_m$, $s_m$ and the melting slope are reported
in Table~\ref{tab:results}.

As a by-product of the simulations on the liquid state I have also
obtained the shear viscosity of the liquid $\eta$, calculated using
the Green-Kubo relations as described in Ref.~\cite{alfe98}, this is
also reported in Table~\ref{tab:results}. Note that the value of the
shear viscosity on the melting curve increases only slightly as a
function of pressure.

In order to test the effect of the size of the simulation cell on the
melting curve, I have also performed one simulation with 1024 atoms
($4\times 4\times 8$ cubic supercell) at $p \sim 47 $~GPa for 11
ps~\cite{time}.
The melting point $(p,T)$ extracted from this simulation is also
reported in Fig.~\ref{fig:melting}, and is essentially
indistinguishable from that obtained with the 432-atom cell.

Other systematic sources of errors are associated with the XC
functional. In order to test how different approximations for the XC
behave, I have used the generalised gradient corrections (GGA)'s known
as PW91~\cite{pw91} and PBE~\cite{pbe}, and evaluated GGA - LDA
differences in the melting temperature at a number of different
pressures.  To do so, I have calculated the free energy differences
between LDA and the GGA functionals employed, for both solid and
liquid, following the techniques described in Ref.~\cite{alfe02b}.  At
zero pressure the difference in free energy between LDA and PW91 is
0.1 eV/atom, and the difference between LDA and PBE is essentially the
same.  When combined with the entropy of melting of $\sim 2.2~k_{\rm
B}$/atom, this difference results in a lowering of the melting
temperature of $\sim 540$~K.  This result is in fair agreement with
the findings of Tangey and Scandolo~\cite{tangney04}, who also
reported that at zero pressure the GGA melting temperature is lower
than the LDA by $\sim 450$~K.  So, similarly to the case of
Al~\cite{alfe03,dewijs98,vocadlo02}, in MgO the GGA underestimates
significantly the zero pressure melting temperature.  However, this
result should not be taken as a general trend of GGA versus LDA, as in
Si, for example, the situation is reversed, with the LDA the zero
pressure melting temperature being lower than the GGA
one~\cite{sugino95,alfe03a}.
At higher pressures the free energy differences between GGA and LDA
are greatly reduced, and at 135 GPa the GGA melting temperature is
lower than the LDA one by only $\sim 100$~K.

Finally, I have tested the effect of the choice of the distribution
between valence and core electrons in the Mg potential, by evaluating
the free energy differences between the MgO system represented with
the couple of Mg and O PAW potentials described above and the system
in which the Mg PAW potential was chosen to have both $3s^2$ and
$2p^6$ in valence. The free energy differences are extremely small,
both at zero and at high pressure, and result in corrections to the
melting temperature of $26$~K and $69$~K at 0 and 135 GPa
respectively. These corrections have been included in the results
reported in Fig.~\ref{fig:melting}

The present results for the melting curve of MgO, based on first
principle simulations of direct coexistence of solid and liquid,
support the ``high'' melting curve previously indicated by a number of
theoretical approaches based on empirical
potentials~\cite{vocadlo96,cohen94,belonoshko96,strachan01}, but are
at variance with the experimental results of ZB. The agreement between
the LDA and GGA predictions for the high pressure melting curve of MgO
suggests that the expected systematic error due to the XC functional
employed may be small. However, this is not the case in the low
pressure region, where the difference between LDA and GGA is of the
order of $\sim 18 \%$.  This relatively large discrepancy points
towards the need of going beyond DFT with the current implementations
of the XC functionals. We believe that one possible way forward will
be the use of quantum Monte Carlo techniques~\cite{foulkes01}.

The predicted high melting curve of MgO has important consequences for
our understanding of the Earth's lower mantle and the history of the
Earth's formation. The melting temperature of the mantle is
constrained between the melting temperatures of the end members
Mg(Fe)O-magnesiow\"ustite and Mg(Fe)SiO$_3$-perovskite, and the
eutectic point. The very high melting temperatures of both MgO and
MgSiO$_3$ (between 7000 and 8500~K~\cite{zerr93}) indicate that the
eutectic temperature of the lower mantle is much higher than its
current temperature (estimated to be between 2550 and 2750 K) and than
the temperature at the top of the core ( $\sim 4000$~K ). This would
suggest that the presence of partial melt of Mg-bearing phase in the
ultra-low velocity zone at the bottom of the Earth's mantle is
unlikely.

This work has been supported by the Royal Society and by the
Leverhulme Trust. The author wishes to acknowledge the computational
facilities of the UCL HiPerSPACE Centre and the national HPC{\em x}
and CSAR national computer facilities granted by NERC through support
of the Mineral Physics Consortium, and by EPSRC through support of the
U.K.C.P. Consortium. Calculations have also been performed on the
Altix machine at University College London provided by the SRIF
programme. The author is grateful to C. Wright for computational
technical support and to M. J. Gillan, G. D. Price and D. Dobson for
very useful discussions.

\twocolumn[\hsize\textwidth\columnwidth\hsize\csname@twocolumnfalse\endcsname
\begin{table}
\begin{tabular}{lccc}
          & $a_0$(\AA) & $B_0$(GPa) & $P_{\rm tr}$(B1-B2) (GPa)\\
\hline
Experiments & 4.213$^a$  4.211$^b$ 4.212$^c$ 4.19$^d$ &  160$\pm 2^a$ 160.2$^c$ 156$^e$ 164.6$^d$ & $>$ 227$^f$   \\
DFT-LDA (large core) & 4.151(4.180) & 180(170) &  503   \\
DFT-LDA (small core) & 4.165(4.194) & 177(167) &  505   \\
DFT-GGA (PW91) & 4.234(4.263) & 158(148) & 491  \\
\end{tabular}
\caption{Experimental and calculated lattice parameter $a_0$ and bulk
modulus $B_0$ of MgO in the NaCl structure (B1), and transition
pressure $P_{\rm tr}$(B1-B2) between the NaCl and the CsCl (B2)
structures. Values in parenthesis include zero point motion and room
temperature effects estimated from Ref.~\protect\cite{karki00}. The
calculations labelled ``large core'' have been performed with the Mg
potential with only the $3s^2$ electrons in valence, while those
labelled ``small core'' with the Mg potential with both the $3s^2$ and
the $2p^6$ electrons in valence.}\label{tab:struct}
\end{table}

\begin{table}
\begin{tabular}{lcccccc}
$p_m$~(GPa) & $T_m^{\rm LDA}$~(K) & $T_m^{\rm GGA}$~(K) & $dT/dp$~(K/GPa) & $v_m$~(\AA$^3$/atom)& $s_m$~($k_{\rm B}$/atom) & $\eta$~(mPa~s)\\
\hline
-0.4(2)  & 3070(50) & 2533(100) & 102(5) & 3.08(5) & 2.19(10) & 3.0(4) \\ 
17.0(2)  & 4590(50) & 4405(100) & 62(3)  & 1.44(5) & 1.69(4) & 3.8(3) \\
47.0(2)  & 6047(40) & 5887(90)  & 33(2)  & 0.73(3) & 1.59(3) & 4.5(4) \\
135.6(2) & 8144(40) & 8044(80)  & 16(1)  & 0.34(2) & 1.51(2) & 5.0(4) \\ 
\end{tabular}
\caption{Calculated melting properties of MgO: pressure $p_m$, LDA and
GGA melting temperatures $T_m^{\rm LDA}$ and $T_m^{\rm GGA}$, slope of
the melting curve $dT/dp$, volume and entropy change on melting $v_m$
and $s_m$, and shear viscosity of the liquid $\eta$ at the melting
point. All quantities (except $T_m^{\rm GGA}$) have been calculated
using DFT-LDA. }\label{tab:results}
\end{table}
]


\begin{thebibliography}{99}
\bibitem{velocity} Initial velocities are assigned according to a
Maxwellian distribution.
\end{thebibliography}
\end{document}